\documentclass[preprint,12pt, a4paper]{elsarticle}

\usepackage{amssymb}
\usepackage{amsmath}

\usepackage{lineno}

\usepackage{float}
\restylefloat{table}

\journal{SoftwareX}

\begin{document}

\begin{frontmatter}



\title{Efficient Bayesian generalized linear models with time-varying coefficients: The walker package in R}


\author{Jouni Helske}

\address{Department of Mathematics and Statistics, University of Jyvaskyla, FI-40014 Jyv{\"a}skyl{\"a}, Finland}
\ead{jouni.helske@jyu.fi}

\begin{abstract}
The R package walker extends standard Bayesian general linear models to the case where the effects of the explanatory variables can vary in time. This allows, for example, to model the effects of interventions such as changes in tax policy which gradually increases their effect over time. The Markov chain Monte Carlo algorithms powering the Bayesian inference are based on Hamiltonian Monte Carlo provided by Stan software, using a state space representation of the model to marginalise over the regression coefficients for efficient low-dimensional sampling.
\end{abstract}


\end{frontmatter}

\section*{Required Metadata}
\label{sec:meta}

\section*{Current code version}
\label{sec:currentcode}
\begin{table}[H]
\begin{tabular}{|l|p{6.5cm}|p{6.5cm}|}
\hline
\textbf{Nr.} & \textbf{Code metadata description} & \textbf{Please fill in this column} \\
\hline
C1 & Current code version & 0.4.1 \\
\hline
C2 & Permanent link to code/repository used for this code version & https://github.com/helske/walker\\
\hline
C3 & Code Ocean compute capsule & none \\
\hline
C4 & Legal Code License   & GPL3 \\
\hline
C5 & Code versioning system used & git \\
\hline
C6 & Software code languages, tools, and services used & R, Stan, C++. \\
\hline
C7 & Compilation requirements, operating environments \& dependencies & R version 3.4.0 and up, C++14, R packages bayesplot, BH, coda, dplyr, ggplot2, Hmisc, KFAS, methods, rlang, rstan, rstantools, StanHeaders, Rcpp, RcppArmadillo, RcppEigen
\\
\hline
C8 & If available Link to developer documentation/manual & https://cran.r-project.org/web/packages/walker/walker.pdf \\
\hline
C9 & Support email for questions & jouni.helske@iki.fi\\
\hline
\end{tabular}
\caption{Code metadata}
\label{metatable} 
\end{table}


\section{Motivation and significance}
\label{}

Assume a time series of interest $y_1,\ldots, y_T$ which is linearly dependent on the some other predictor time series $X_1,\ldots,X_T$, where $X'_t = (x_{1t},\ldots,x_{pt})'$ are the predictor variables at the time point $t$, and define linear-gaussian time series regression model as
\begin{equation}
\label{eq:tsreg}
\begin{aligned}
y_t &= X'_t\beta_t + \epsilon_t, \quad  t = 1,\ldots, T,
\end{aligned}
\end{equation}
where $\epsilon_t \sim N(0, \sigma_{\epsilon}^2)$ and $\beta$ is a vector of $p$ unknown regression coefficients (first one typically being the intercept term).

It is not always reasonable to assume that the relationship between $y_t$ and some predictor $x_{it}$ stays constant over $t=1,\ldots,T$, the time period of interest. The convenient linear relationship approximation may hold well for only piecewise if the underlying relationship is nonlinear or when there are some unmeasured confounders alter the relationship between the measured variables. Allowing the regression coefficients to vary over time can in some instances alleviate these problems. It is also possible that  our knowledge of the phenomena of interest already leads to suspect time-varying relationships (see, e.g., Chapter 1 in \cite{moryson}).

The basic time series regression model \eqref{eq:tsreg} can be extended to allow the unknown regression coefficients $\beta_t$ to vary over time. This can be done in various ways, for example, by constructing a semiparametric model based on kernel smoothing \cite{robinson}, or parametrically by using dynamic Bayesian networks \cite{Dagum} or state space models \cite{harvey1982}. We follow the state space modelling approach and define a gaussian time series regression model with random walk coefficients as
\begin{equation}
\label{eq:rwtsreg}
\begin{aligned}
y_t &= X'_t \beta_t + \epsilon_t,\\
\beta_{t+1} &= \beta_t + \eta_t,
\end{aligned}
\end{equation}
where $\eta_t \sim N(0, D)$, with $D$ being $p \times p$ diagonal matrix with diagonal elements $\sigma^2_{i,\eta}$, $i=1,\ldots,p$, and define a prior distribution for the first time point $\beta_1$ as $N(\mu_{\beta_1}, \sigma^2_{\beta_1})$. The bottom equation in \ref{eq:rwtsreg} defines a random walk process for the regression coefficients with $D$ defining the degree of variability of the coefficients (with $D=0$ the model collapses to basic regression model). 

Our goal is a Bayesian estimation of the unknown regression coefficients $\beta_1,\ldots, \beta_T$ and standard deviations $\sigma = (\sigma_{\epsilon}, \sigma_{1, \eta}, \ldots, \sigma_{p, \eta})$. Although in principle we can estimate these using the general-purpose Markov chain Monte Carlo (MCMC) software such as \verb|Stan| \cite{stan} or \verb|BUGS| \cite{lunn-thomas-best-spiegelhalter}, these standard implementations can be computationally inefficient and prone to severe problems related to the convergence of the underlying MCMC algorithm due to the nature of our models of interest. For example in (block) Gibbs sampling approach we target the joint posterior $p(\beta, \sigma | y)$ by sampling from $p(\beta | \sigma, y)$ and $p(\sigma | \beta, y)$. But because of strong autocorrelations between the coefficients $\beta$ at different time points, as well as with the associated standard deviation parameters, this MCMC scheme can lead to slow mixing. Also, the total number of parameters to sample increases with the number of data points $T$. Although the Hamiltonian Monte Carlo algorithms offered by \verb|Stan|  are typically more efficient exploring high-dimensional posteriors than Gibbs-type algorithms, we still encounter similar problems.

An alternative solution used by the \verb|R| \cite{R} package \verb|walker| is based on the property that model \eqref{eq:rwtsreg} can be written as a linear gaussian state space model (see, e.g., Section 3.6.1 in \cite{DK2012}). This allows us to marginalize the regression coefficients $\beta$ during the MCMC sampling by using the Kalman filter leading to a fast and accurate inference of marginal posterior $p(\sigma | y)$. Then, the corresponding joint posterior $p(\sigma, \beta | y) = p(\beta | \sigma, y)p(\sigma | y)$ can be obtained by simulating the regression coefficients given marginal posterior of standard deviations. This sampling can be performed for example by simulation smoothing algorithm \cite{durbin-koopman2002}.

The marginalization of regression coefficients cannot be directly extended to generalized linear models such as Poisson regression, as the marginal log-likelihood is intractable. However, it is possible to use Gaussian approximation of this exponential state space model \cite{durbin-koopman1997}, and the resulting samples from the approximating posterior can then be weighted using the importance sampling type correction \cite{vihola-helske-franks}, leading again to asymptotically exact inference.

When modelling the regression coefficients as a simple random walk, the posterior estimates of these coefficients can have large short-term variation which might not be realistic in practice. One way of imposing more smoothness for the estimates is to switch from random walk coefficients to integrated second order random walk coefficients, defined as
\begin{equation*}
\begin{aligned}
\beta_{t+1} &= \beta_t + \nu_t,\\
\nu_{t+1} &= \nu_t + \xi_t,
\end{aligned}
\end{equation*}

with $\xi_t \sim N(0, D_{\xi})$. This is a local linear trend model \cite{harvey} (also known as an integrated random walk), with the restriction that there is no noise on the $\beta$ level. For this model, we can apply the same estimation techniques as for the random walk coefficient case. More complex patterns for $\beta$ are also possible. For example, we can define that $\beta_{t+1} \sim N(\beta_t,\gamma_t \sigma_{\eta})$ where $\gamma_1,\ldots,\gamma_T$ is known monotonically decreasing sequence of values leading to a case where $\beta_t$ gradually converges to constant over time.

\section{Software description}
\label{sec:desc}

A stable version of \verb|walker| is available at CRAN\footnote{https://cran.r-project.org/package=walker}, while the current development version can be installed from Github\footnote{https://github.com/helske/walker}. The MCMC sampling is handled by \verb|rstan| \cite{rstan}, the \verb|R| interface to \verb|Stan|, while the model definitions and analysis of the results are performed in \verb|R|, leading to fast and flexible modelling.

For defining the models, the \verb|walker| package uses similar Wilkinson-Rogers model formulation syntax \cite{Wilkinson1973} as, for example, basic linear model function \verb|lm| in \verb|R|, but the formula for \verb|walker| also recognizes two custom functions, \verb|rw1| and \verb|rw2| for random walk and integrated random walk respectively. For example, by typing
\begin{verbatim}
fit <- walker(y ~ 0 + x + 
  rw1(~ z, beta = c(0, 1), sigma = c(0, 1)), 
  sigma_y = c(0, 1))
\end{verbatim}
we define a model with time-invariant coefficient for predictor $x$, and first order random walk coefficients for $z$ and time-varying intercept term. Priors for $\beta$ (including the intercept) and $\sigma$ are defined as vectors of length two which define the mean and standard deviation of the normal distribution respectively (for standard deviation parameters the prior is truncated at zero).

Function \verb|walker| creates the model based on the formula and the prior definitions, and then calls the \verb|sampling| function from the \verb|rstan| package. The resulting posterior samples can be then converted to a data frame format using the function \verb|as.data.frame| for easy visualization and further analysis.

In addition to the main functions \verb|walker| for the Gaussian case and \verb|walker_glm| for the Poisson and negative binomial models, the package contains additional functions for visualization of the results (e.g., \verb|plot_coef| and \verb|pp_check|) and out-of-sample prediction (function \verb|predict|). Also, as the modelling functions return the full \verb|stanfit| used in MCMC sampling, this object can be analysed using many general diagnostic and graphical tools provided by several \verb|Stan| related \verb|R| packages such as \verb|ShinyStan| \cite{shinystan}.

\section{Illustrative Examples}
\label{sec:examples}

As an illustrative example, let us consider a observations $y$ of length $n=100$, generated by random walk (i.e. time varying intercept) and two predictors. First we simulate the coefficients, predictors and the observations:
\begin{verbatim}
set.seed(1)
n <- 100
beta1 <- cumsum(c(0.5, rnorm(n - 1, 0, sd = 0.05)))
beta2 <- cumsum(c(-1, rnorm(n - 1, 0, sd = 0.15)))
x1 <- rnorm(n, mean = 2)
x2 <- cos(1:n)
rw <- cumsum(rnorm(n, 0, 0.5))
signal <- rw + beta1 * x1 + beta2 * x2
y <- rnorm(n, signal, 0.5)
\end{verbatim}

Then we can call function \verb|walker|. As noted in Section \ref{sec:desc} the model is defined as a simple formula object, and in addition to the prior definitions we can pass various arguments to the \verb|sampling| method of \verb|rstan|, such as the number of iterations \verb|iter| and the number of chains \verb|chains| to be used for the MCMC (default values for these are 2000 and 4 respectively). 

\begin{verbatim}
fit <- walker(y ~ 0 + rw1(~ x1 + x2, 
  beta = c(0, 10), sigma = c(0, 10)), 
  sigma_y = c(0, 10), chains = 2, seed = 1)
\end{verbatim}

The output of \verb|walker| is \verb|walker_fit| object, which is essentially a list with \verb|stanfit| from \verb|Stan|'s \verb|sampling| function, the original observations \verb|y| and the covariate matrix \verb|xreg|. This allows us to use all the postprocessing functions for \verb|stanfit| objects.

\begin{figure}[!h]
	\centering
	\includegraphics[width=\textwidth]{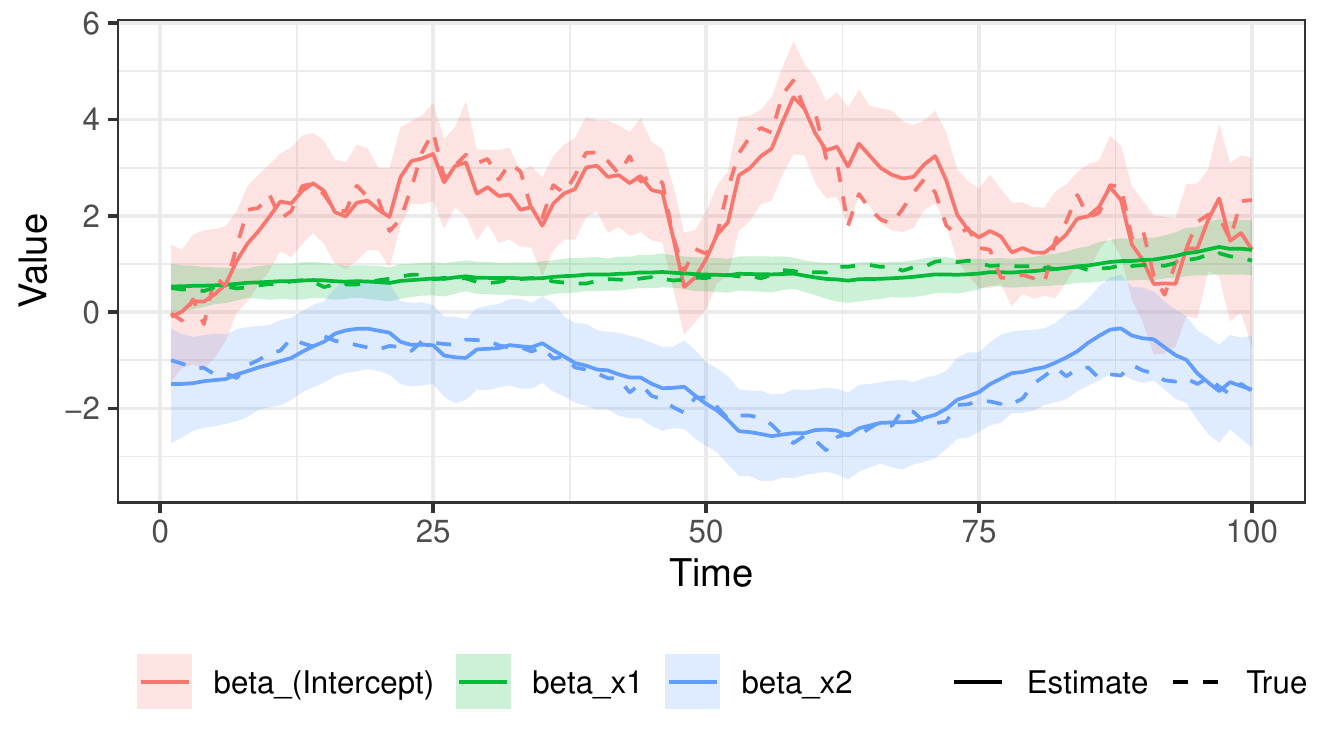} 
	\caption{Posterior means (solid lines) and 95\% posterior intervals (shaded areas) of the time-varying regression coefficients and the true data generating values (dashed lines) of the illustrative example.}
	\label{fig:betas}
\end{figure}

Figure \ref{fig:betas} shows how \verb|walker| recovers the true coefficient processes relatively well: The posterior intervals contain the true coefficients and the posterior mean estimates closely follow the true values. For drawing the figure, we first use function \verb|as.data.frame| to extract our posterior samples of the coefficients as data frame, then use packages \verb|dplyr| \cite{dplyr} and \verb|ggplot2| \cite{ggplot2} to summarise and plot the posterior means and 95\% posterior intervals respectively: 

\begin{verbatim}
library(dplyr)
library(ggplot2)
sumr <- as.data.frame(fit, "tv") %>% 
  group_by(variable, time) %>%
  summarise(mean = mean(value), lwr = quantile(value, 0.025),
    upr = quantile(value, 0.975))
sumr$true <- c(rw, beta1, beta2)

ggplot(sumr, aes(y = mean, x = time, colour = variable)) + 
  geom_ribbon(aes(ymin = lwr, ymax = upr, fill = variable), 
    colour = NA, alpha = 0.2) + 
  geom_line(aes(linetype = "Estimate"), lwd = 1) + 
  geom_line(aes(y = true, linetype = "True"), lwd = 1) +
  scale_linetype_manual(values = c("solid", "dashed")) +
  theme_bw() + xlab("Time") + ylab("Value") +
  theme(legend.position = "bottom", 
    legend.title = element_blank())
\end{verbatim}

More examples can be found in the package vignette and function documentation pages (e.g. typing \verb|vignette("walker")| or \verb|?walker_glm| in \verb|R|), including a comparison between the marginalization approach of \verb|walker| and "naive" implementation with \verb|Stan|, and an example on the scalability.

\section{Impact and conclusions}
\label{impact}

The \verb|walker| package extends standard Bayesian generalized linear models to flexible time-varying coefficients case in a computationally efficient manner, which allows researchers in economics, social sciences and other fields to relax the sometimes unreasonable assumption of a stable, time-invariant relationship between the response variable and (some of) the predictors. Similar methods have been previously used in maximum likelihood setting for example in studying the diminishing effects of advertising \cite{Kinnucan} and demand for international reserves \cite{Bahmani-Oskooee}.

There are several ways how \verb|walker| can be extended in the future. There are already some plans for additional forms of time-varying coefficients (such as a stationary autoregressive process), support for more priors and additional distributions for the response variables (e.g., negative binomial).

\section{Conflict of Interest}

We wish to confirm that there are no known conflicts of interest associated with this publication and there has been no significant financial support for this work that could have influenced its outcome.

\section*{Acknowledgements}
\label{thanks}

This work has been supported by the Academy of Finland research grants 284513, 312605, and 311877.

\bibliographystyle{elsarticle-num} 
\bibliography{walker_revision}

\end{document}